\documentclass[nofootinbib,aps,10pt,twocolumn]{revtex4-1}

\usepackage{amsmath,amssymb,bm,epsfig}
\usepackage{color}
\usepackage{hyperref} 
\usepackage{graphicx}
\usepackage{xcolor,colortbl}
\RequirePackage{lineno}
\definecolor{Gray}{gray}{0.85}
\newcolumntype{a}{>{\columncolor{Gray}}c}

\def \beq{\begin{equation}}
\def \eeq{\end{equation}}
\def \beqa{\begin{eqnarray}}
\def \eeqa{\end{eqnarray}}

\def \l{\left(}
\def \r{\right)}
\def \l{\left(}
\def \r{\right)}

\begin{document}

\title{Splitting of proton-antiproton directed flow in relativistic heavy-ion collisions}

\author{Piotr Bo{\.z}ek}
\email{piotr.bozek@fis.agh.edu.pl}
\affiliation{AGH University of Science and Technology,\\ 
Faculty of Physics and Applied Computer Science,\\
aleja Mickiewicza 30, 30-059 Krakow, Poland}

\begin{abstract}

  The rapidity dependent directed flow of particles produced in a relativistic heavy ion collision can be generated in the hydrodynamic expansion
  of a tilted source.  The asymmetry of the pressure leads to a build up of
  a directed flow of matter with respect to the collision axis.
  The experimentally observed ordering of the directed flow of baryons,
  pions and antibaryons can be described  as resulting from the expansion
  of a  baryon inhomogeneous fireball.  An uneven distribution of baryons
  in the transverse plane leads to an asymmetry in the collective
  push for protons
  and antiprotons. Precise measurements  of the collective flow as a function of rapidity could serve as a strong constraint on mechanism of baryon stopping in the early phase of the collision.

\end{abstract}

\maketitle

The creation of a dense, strongly interacting matter in relativistic
heavy ion collisions manifests itself through the build up of
collective flow of emitted particles \cite{Ollitrault:2010tn,Heinz:2013th,Gale:2013da}.
The collective flow reflects the gradient
of the pressure in the fireball. The collective flow is parameterized using the
harmonic flow coefficient for the azimuthal angle dependence
of the observed spectra of produced particles. In particular, the directed flow with
respect to the reaction plane  $\Psi_R$ is defined as a rapidity dependent  first harmonic coefficient $v_1(y)$
\begin{equation}
  \frac{dN}{d\phi dy} = \frac{dN}{dy} \left[ 1 + 2 v_1(y) \cos(\phi -\Psi_R)+ \dots \right] \ . 
  \end{equation}
In collisions of symmetric nuclei the directed flow is an odd
function of rapidity. 

The observation of the rapidity odd directed flow $v_1$ of charged particles produced in symmetric heavy ion  collisions~\cite{Wetzler:2002fi,Back:2005pc,STAR:2011hyh,Abelev:2008jga}  shows that the forward-backward symmetry is broken.
In dynamic models of collisions it  is due to an  asymmetry of the initial conditions \cite{Csernai:1999nf,Snellings:1999bt,Lisa:2000ip,Adil:2005qn,Bozek:2010bi,Chen:2009xc,Csernai:2011gg,Steinheimer:2014pfa,vanderSchee:2015rta,Becattini:2015ska,Konchakovski:2014gda,vanderSchee:2015rta,Ke:2016jrd,Jiang:2021ajc}. In non-central collisions the initial state 
is expected to break the forward-backward symmetry.
In the hydrodynamic model the directed flow can be explained
by the expansion of a fireball tilted with respect to
the collision axis \cite{Bozek:2010bi}. This phenomenological
model can describe the  measured directed flow of charged particles.
However, it cannot explain the observed splitting of the directed flow of identified particles
\cite{STAR:2011hyh,STAR:2014clz}. In particular, the measured directed flow
is different for protons and antiprotons, while in the hydrodynamic model
particles of the same mass are expected to have a similar collective flow.
Dynamical or hybrid models of heavy-ion collisions cannot explain the observed directed flow for identified particles at different energies \cite{Chen:2009xc,Guo:2012qi,Ivanov:2014ioa,Steinheimer:2014pfa,Konchakovski:2014gda,Nara:2016phs,Guo:2017mkf}.
This long standing problem can be resolved assuming that the baryon distribution in the fireball is inhomogeneous. The baryon chemical potential is not only rapidity dependent \cite{Biedron:2006vf}, but also depends  on the position
in the transverse plane. The correlation between the position of
the stopped baryon in the transverse plane
and the position of the participant nucleon leads
to a tilt of the baryon distribution. In the following,
I show how this mechanism
can explain the observed hierarchy of the directed flows of pions, protons,
and antiprotons.

Dynamical models of the initial state are developed for the description of
the creation and subsequent hydrodynamic
evolution of a three-dimensional fireball in heavy ion collisions
\cite{Shen:2017bsr,Akamatsu:2018olk,Du:2018mpf,Shen:2022oyg,De:2022yxq}.
In this paper, the difference of the average
flow of baryons, antibaryons, and pions is  discussed. The effect on
the average flow can 
be understood using a simple model with smooth initial conditions.
The initial conditions are chosen in the form of  a tilted fireball in the Glauber model \cite{Bozek:2010bi}. The participant densities  in the transverse plane  $(x,y)$ for the  right (+) and left (-) going   nuclei colliding at impact parameter b are
\begin{equation}
  T_{\pm}(x,y)= A \ T(x\pm b/2,y)\left[1 - \left(1- \sigma T(x\mp b/2,y)\right)^A \right] \ , 
\end{equation}
where
\begin{equation}
  T(x,y)= \int dz \rho(\sqrt{x^2+y^2+z^2}) \ ,
  \end{equation}
$\rho$ is the normalized nuclear density profile
\begin{equation}
  \rho(r)=\frac{C}{1+\exp((r-R)/a)} , 
  \end{equation}
$\int d^3r \rho(r)=1$. For the case studied in this paper, Au+Au collisions at $\sqrt{s_{NN}}=200$GeV, the parameters are $A=197$, $\sigma=42$~mb, $R=6.37$~fm,
$a=0.54$~fm.
The number of participant nucleons at impact parameter b  is $N_{part}=\int dx dy \left[ T_{+}(x,y)+ T_{-}(x,y) \right]$. The average multiplicity of produced particles is proportional to a combination
$(1-\alpha)N_{part}+2 \alpha N_{coll}$, where the number of binary collisions
$N_{coll}=\sigma A^2  \int dx dy T(x+b/2,y) T(x+b/2,y)$.
The three dimensional distribution of the initial entropy density in the transverse plane and space-time rapidity $\eta_{||}=\frac{1}{2}\ln\left(\frac{t+z}{t-z}\right)$ is
\begin{eqnarray} \nonumber
 & & s(x,y,\eta_S) \propto    \left[ (1-\alpha) \left(T_{+}(x,y)f_{+}(\eta_s) +T_{-}(x,y)f_{-}(\eta_s) \right)  \right. \\
    & & \left. + 2 \alpha \sigma A^2  T(x+b/2,y) T(x+b/2,y)\right] H(\eta_S)  \ .
  \label{eq:sdensity}
\end{eqnarray}
The two terms in the density correspond to the participant and binary collisions contributions to the overall multiplicity of produced particles.
The longitudinal profile is
\begin{equation}
  H(\eta_S)= \exp\l-\theta\l|\eta_S|-\eta_S^0\r\frac{\l|\eta_S|-\eta^0_S\r^2}{2\sigma_\eta^2}\r \ , 
  \end{equation}
with $\eta_S^0=1$ and $\sigma_\eta=1.3$. The average  entropy deposited in the interaction region  by a participant nucleon is asymmetric in the
longitudinal direction ~\cite{Brodsky:1977de,Bialas:2004su,Adil:2005qn}. The
corresponding profiles  for left  and right going participants are
\begin{equation}
  f_{\pm}(\eta_{||})= \left\{\begin{array}{lr}
                            1, & \eta_{||}>\eta_T\\
                            \frac{\eta_T \pm \eta_{||}}{2\eta_T}, & -\eta_T\leq\eta_{||}\leq\eta_T\\
                            0, & \eta_{||}<-\eta_T
                           \end{array}\right. \ \ .
\label{eq:fpm}
 \end{equation}
The linear form of the  asymmetric profiles $f_{\pm}$   can describe the tilt of the fireball in the central rapidity region. Other ans{\"a}tze
  used \cite{Jiang:2021ajc,Jiang:2022uoe} for the tilted fireball are very similar
  in this  central rapidity region. 
The tilted fireball scenario  has been successful is describing the observed charged particle 
directed flow~\cite{Bozek:2010bi}.
The component of the initial density corresponding
to  the binary collision contribution is not tilted. 
The tilt of the fireball in the central region is
determined by two parameters  $\eta_T=2.4$ and $\alpha=0.1$.
Please note that the asymmetric entropy deposition is an average description of
of fluctuating string initial conditions 
\cite{Bozek:2015bna,Shen:2017bsr}.   The tilted and non-tilted component  reflects qualitatively the possibility of different kind of strings formed in the initial state, involving valence and see quarks from the  participant nucleons.

The hydrodynamic expansion is performed using the three-dimensional
version of the hydrodynamic code MUSIC \cite{Schenke:2010nt,Schenke:2010rr,Paquet:2015lta}
with shear viscosity $\eta/s=0.08$ and an equation of state for quark-gluon plasma at finite baryon density \cite{Monnai:2019hkn}. The freeze-out takes place at the density $\epsilon_{fr}=0.5$~GeV/fm$^3$. In the very early stage of the collision a prequilibrium expansion is usually assumed. However, the existing implementations of the prequilibrium expansion use a two-dimensional, boost-invariant geometry \cite{Kurkela:2018vqr}. Whereas, the early stage of the build up of the directed flow is very sensitive to the relative values of the transverse and longitudinal effective pressures in the fluid \cite{Bozek:2010aj}. The investigation of the 
 details of such a  very early three-dimensional evolution is beyond the scope of the present phenomenological study. Instead, a
viscous hydrodynamic evolution is imposed from an early initialization 
time $0.2$~fm/c.  

\begin{figure}
 \begin{center}
   \includegraphics[scale=0.48]{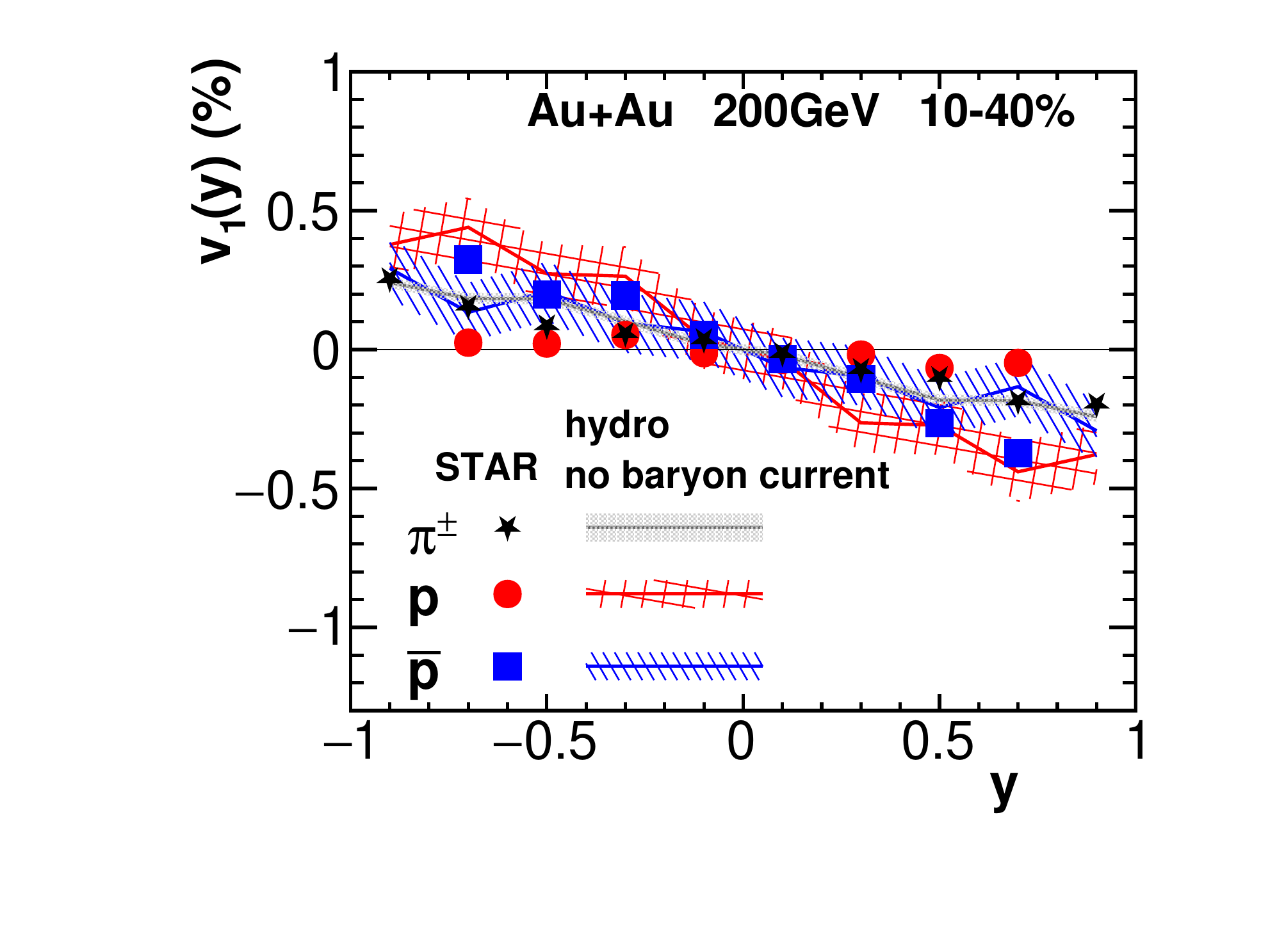}
   \vskip -10mm
 \caption{ Rapidity dependence of the directed flow of pions (stars and gray band), protons (full circles and red band) and antiprotons (squares and blue band)
 in $10-40$\% centrality  Au+Au collisions. The symbols represent the STAR Collaboration data \cite{STAR:2014clz} and the bands the results of the hydrodynamic expansion of a  baryon homogeneous fireball.}
 \label{fig:v1etanb}
 \end{center}
\end{figure}

The expansion of the tilted fireball generates the rapidity
dependent directed flow. The model results reproduce reasonably well the
directed flow of  charged particles \cite{Bozek:2010bi},
but not the splitting of  directed flows of
protons and anti-protons \cite{STAR:2011hyh,STAR:2014clz}. The same kinematic cuts are used for pions and protons as in the experimental analysis.
The
directed flow of pions as a function of rapidity can be reproduced by the model using a tilted source initial conditions for the bulk of the matter (Fig. \ref{fig:v1etanb}).
In the calculation there is no baryon current and the flow of protons and antiprotons is similar.
This is not surprising, as in this version of the hydrodynamic model
the predicted flow depends only on the particle mass. The model
cannot predict the observed splitting of the directed
flow of baryons and antibaryons. At lower energies the splitting of
baryon and antibaryon flow in hydrodynamic models could be related to a possible phase transition
\cite{Konchakovski:2014gda,Ivanov:2014ioa,Steinheimer:2014pfa,Nara:2016phs}, to an effect of the spectator matter 
 \cite{Zhang:2018wlk}, or to the electromagnetic
effect \cite{Rybicki:2013qla}. At the energy $\sqrt{s_{NN}}=200$~GeV these effect cannot explain the observed splitting of directed flow. The baryon-antibaryon
splitting of the directed flow indicates
that  the baryons transported from the initial colliding nuclei have effectively
a different flow than the produced baryons \cite{STAR:2017okv}. 

\begin{figure}
 \begin{center}
   \includegraphics[scale=0.48]{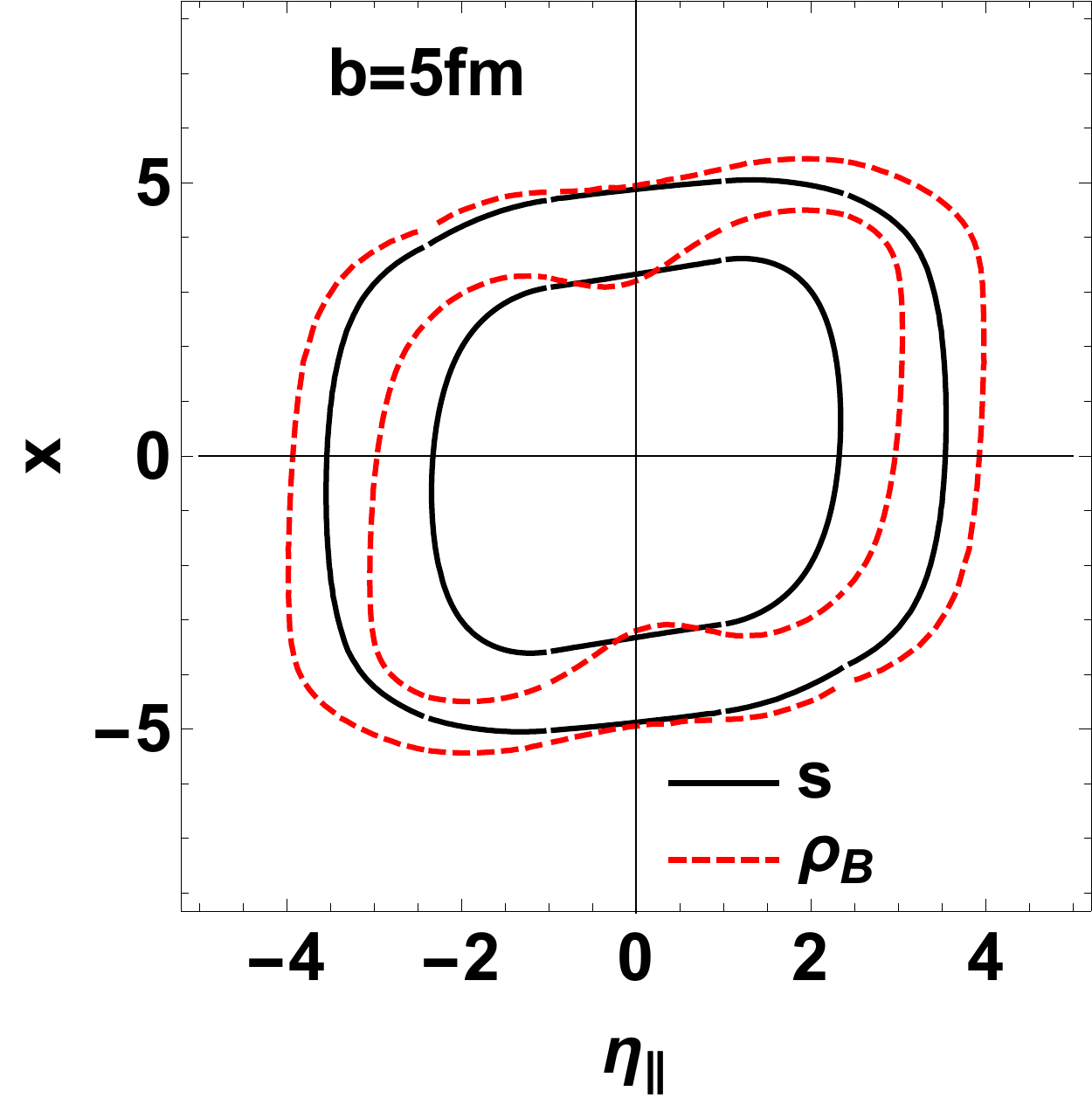}
 %  \vskip -2mm
 \caption{ Initial entropy density $s(x,0,\eta_S)$ (solid contours) and baryon density $\rho_b(x,0,\eta_s)$ (dashed contours). }
 \label{fig:init}
 \end{center}
\end{figure}

In the hydrodynamic model the difference in the  flow of baryons and
anti-baryons indicates
that the net baryon density is finite  and baryon currents
are present in the  expanding
fireball. The initial baryon density is assumed to be of the form
\begin{eqnarray} \nonumber
  \rho_B(x,y,\eta_{||}) & \propto & \left[ \left(T_{+}(x,y)f_{+}(\eta_s) +T_{-}(x,y)f_{-}(\eta_s) \right)\right] \\ & & H(\eta_{||}) H_B(\eta_{||}) \,
  \label{eq:rhob}
\end{eqnarray}
where
$
H_B(\eta_{||})=\exp\left( -\frac{(\eta_{||}-\eta_b)^2}{2\sigma_B^2}\right)+\exp\left( -\frac{(\eta_{||}+\eta_b)^2}{2\sigma_B^2}\right)$, with  parameters
$\eta_B=4.4$ and $\sigma_B=2.2$. The ratio  of the transported and produced baryons at central rapidities \cite{STAR:2008med} is reproduced with these initial conditions, which is essential for the the description of the
proton-antiproton splitting of directed flow.
Please note, that the parametrization (Eq. \ref{eq:rhob}) of the baryon density is tilted with respect to the collision axis, but the tilt is slightly larger than for the bulk of the matter (Eq. \ref{eq:sdensity}). The ansatz used is phenomenological, but qualitatively justified in string models of the initial state \cite{Bialas:2004kt,Shen:2017bsr,Jezabek:2021oxg}.
The net baryon number transported to central rapidities must originate from the valence quarks in the participant nucleons. Therefore, the net baryon density
in the transverse plane is strongly correlated to the density
of participant nucleons. On the other hand, the
deposited entropy is expected to originate from strings involving valence and sea quarks. This  implies that the tilt of the initial entropy density could be
smaller than the tilt of the initial baryon distribution. In Fig. \ref{fig:init} are shown the contour plots for the initial entropy and baryon number density distributions for
a Au+Au collision at $b=5fm$.  The larger tilt of the initial baryon
density generates inhomogeneities
in the proton and antiproton distribution in the fireball.
At positive space-time rapidities the proton density is larger in the
upper part of the fireball on the figure,
the reverse is true for the antiproton density.

\begin{figure}
 \begin{center}
   \includegraphics[scale=0.48]{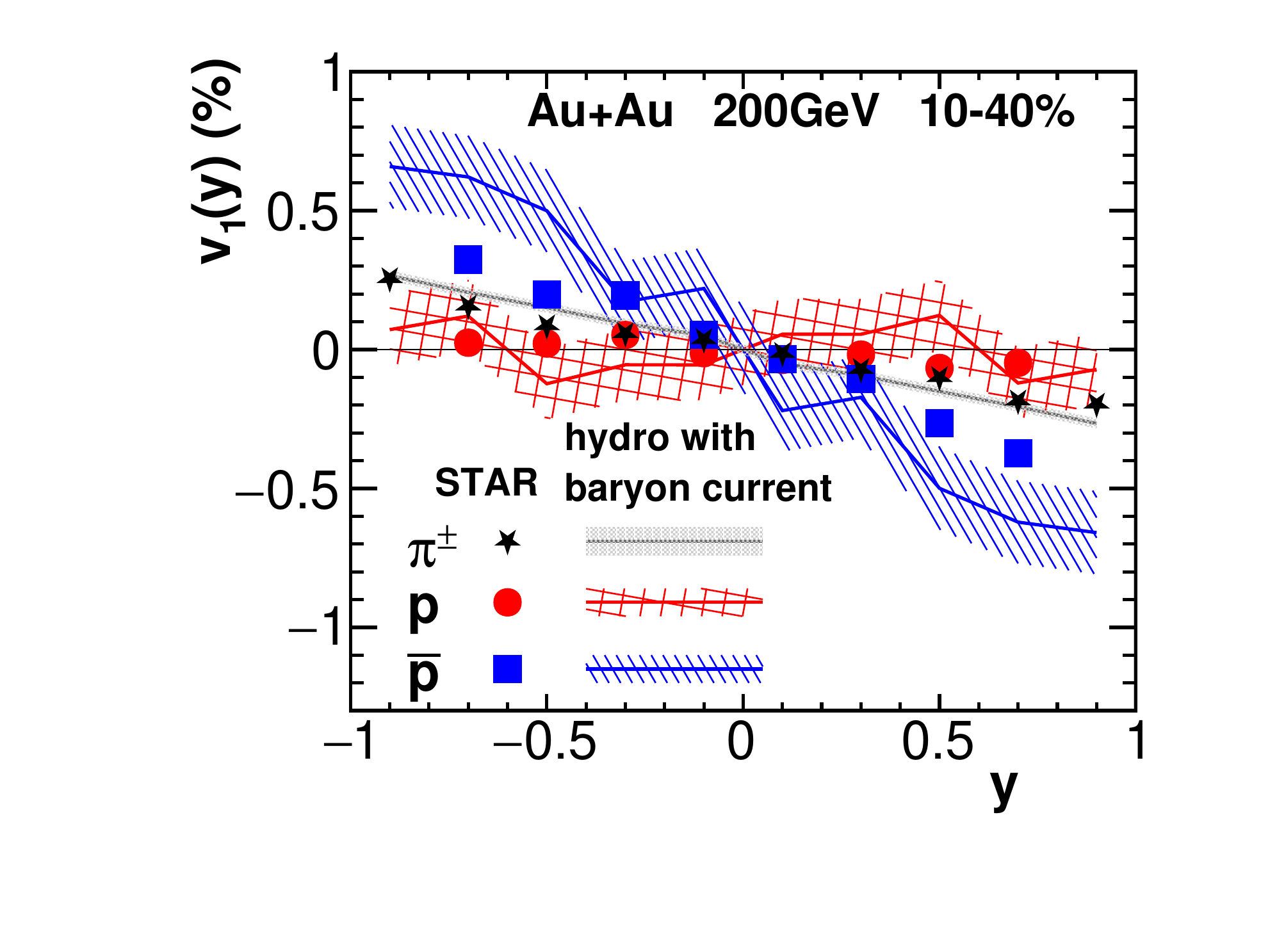}
   \vskip -10mm
 \caption{ Rapidity dependence of the directed flow of pions (stars and gray band), protons (full circles and red band) and antiprotons (squares and blue band)
 in $10-40$\% centrality  Au+Au collisions. The symbols represent the STAR Collaboration data \cite{STAR:2014clz} and the bands the results of the hydrodynamic calculation with baryon current.}
 \label{fig:v1eta}
 \end{center}
\end{figure}

The initial energy and baryon distributions are propagated using
a viscous hydrodynamic evolution with baryon current using the MUSIC code.
At the freeze-out particles are emitted  following the local temperature,
flow velocity, and baryon chemical potential in a fluid cell, as well as
shear viscosity corrections.
As a result, the final proton directed flow of protons (antiprotons)
is larger (smaller) than directed flow of pions at positive rapidities (Fig. \ref{fig:v1eta}). The model calculation reproduces well the rapidity dependent directed flow for pions, similarly as for the calculation without baryon current (Fig. \ref{fig:v1etanb}).

\begin{figure}
 \begin{center}
   \includegraphics[scale=0.48]{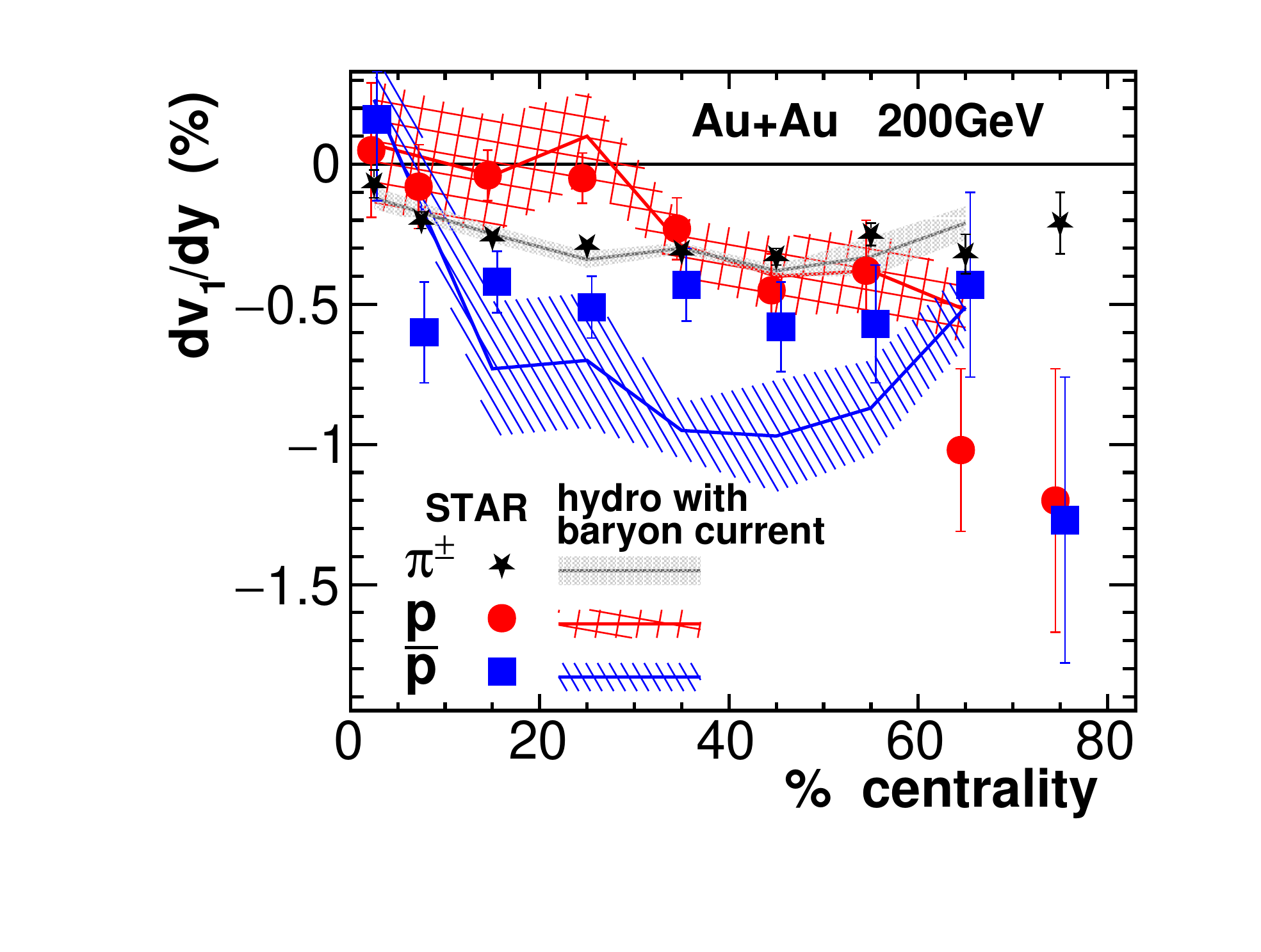}
   \vskip -10mm
 \caption{The slope of the rapidity dependence of the directed flow of pions, protons and antiprotons as function of centrality in Au+Au collisions. Symbols as in Fig. \ref{fig:v1eta}.}
 \label{fig:dv1}
 \end{center}
\end{figure}

The tilt of the bulk of the fireball matter and  the  tilt of the baryon density
change with the centrality of the collision.
The  slope of the rapidity dependence of the directed flow $dv_1(y)/dy$ is fitted in the central rapidity region $y\in [-1,1]$ for each particle species at a range of centralities from $0\%$ to $70\%$. The result as a function of
centrality is shown in Fig. \ref{fig:dv1} and compared to STAR Collaboration data. In semicentral collisions
the model and the experimental data show the largest splitting between
the slopes of the directed flow for the three particle species studied. The
splitting is reduced for semiperipheral collisions in the experiment.
Qualitatively this is also seen in the model calculation.
The calculation is based on a phenomenological ansatz. The same phenomenological  parameters  determining the tilt of the fireball are used
at all centralities and the experimental
directed flow slopes for identified particles   at different centralities are
reproduced  only qualitatively.

The analysis
demonstrates the origin of the observed splitting of the baryon-antibaryon
directed flow, but cannot replace microscopic models of the initial state.
The directed flow of identified particles could serve as strong experimental
constraint in the development of such realistic models of the initial state.
The proposed mechanism could qualitatively explain the observed
splitting of the
baryon-antibaryon directed
flow also at lower energies \cite{STAR:2014clz} (with adjusted parameters).
Again, a detailed study
of the energy dependence of the directed flow splitting
should be rather based on more realistic models.
For $\sqrt{s_{NN}}<20$GeV effects due to phase transition,
shadowing by spectators or a mean field could also play a role.

The paper presents a mechanism which generates the splitting of the
directed flow for protons and antiprotons in hydrodynamic models.
This scenario strongly suggests that the
distribution of baryons in the fireball is inhomogeneous in the transverse
plane.  This characteristic of the initial state could serve as
a constraint for dynamical models of the initial state involving
stopped baryons. This effect could have consequences for precise studies
of the collective flow harmonics for identified particles or of the
polarization
for baryons and antibaryons \cite{STAR:2017ckg,Wu:2022mkr}. Similarly, other
conserved charges could  be unevenly distributed in the fireball.
However, for strangeness
the most important effect to be considered could be  the different rate of
strangeness saturation between the core and corona of the fireball and not the
inhomogeneities of the net strangeness. Finally, let us note that a similar experimental and theoretical analysis could be performed for collision of asymmetric nuclei to gain additional information on the initial state of the hydrodynamic evolution.

This research is partly supported by
the National Science Centre Grant No. 2018/29/B/ST2/00244.

\bibliography{../hydr}

\end{document}